\documentclass[twoside,a4paper]{article}
\usepackage{fleqn,epsf,graphicx}
\usepackage{authblk}
\usepackage[utf8]{inputenc}
\usepackage{fancyhdr}

\begin{document}

\title{A pilot study of the novel J-PET plastic scintillator 
with 2-(4-styrylphenyl)benzoxazole as~a~wavelength shifter}

\author{A.~Wieczorek$^{a,}$ $^{b}$, P.~Moskal$^{a}$, S.~Niedźwiecki$^{a}$, T.~Bednarski$^{a}$, P.~Białas$^{a}$, E.~Czerwiński$^{a}$, A.~Danel$^{c}$, A.~Gajos$^{a}$, A.~Gruntowski$^{a}$, D.~Kamińska$^{a}$, Ł.~Kapłon$^{a,}$ $^{b}$, A.~Kochanowski$^{d}$, G.~Korcyl$^{a}$, J.~Kowal$^{a}$, P.~Kowalski$^{e}$, T.~Kozik$^{a}$, W.~Krzemień$^{a}$, E.~Kubicz$^{a}$, M.~Molenda$^{d}$, M.~Pałka$^{a}$, L.~Raczyński$^{e}$, Z.~Rudy$^{a}$, O.~Rundel$^{a}$, P.~Salabura$^{a}$, N.G.~Sharma$^{a}$, M.~Silarski$^{a}$, A.~Słomski$^{a}$, J.~Smyrski$^{a}$, A.~Strzelecki$^{a}$, T.~Uchacz$^{d}$, W.~Wiślicki$^{e}$, M.~Zieliński$^{a}$, N.~Zoń$^{a}$}

\affil{$^{a}$Faculty of Physics, Astronomy and Applied Computer Science, Jagiellonian University\\

       $^{b}$Institute of Metallurgy and Materials Science, Polish Academy of~Sciences
       
       $^{c}$Institute of Chemistry and Physics, University of Agriculture
       
       $^{d}$Faculty of Chemistry, Jagiellonian University
       
       $^{e}$Świerk Computing Centre, National Centre for Nuclear Research}

\maketitle                   


\begin{abstract}
  For the first time a molecule of 2-(4-styrylphenyl)benzoxazole containing benzoxazole and stilbene groups is applied as a scintillator dopant acting as a wavelength shifter. In this article a light yield of the plastic scintillator, prepared from styrene doped with 2 wt\% of 2,5-diphenylbenzoxazole and 0.03 wt\% of 2-(4-styrylphenyl)benzoxazole, is determined to be as large as 60\% $\pm$ 2\% of the anthracene light output. There is a potential to improve this value in the future by the optimization of the additives concentrations. 
\end{abstract}

%
%

\section{Introduction}

Scintillators are widely used as detectors of radiation in many fields of science and industry. For example they are crucial components of apparatus used in nuclear and particle physics experiments, medical diagnostics modalities as well as radiation detectors in many sectors of homeland security \cite{a1}

Scintillators play an important role in most of the nuclear imaging devices such as e.g.  Positron Emission Tomography device (PET). Though PET is well established technology there is still endeavour to improve its performance by e.g. searching for a new kind of crystals with better timing properties \cite{a2,a3,a4,a5}.  All of the currently available PET scanners are based on the inorganic crystals. However, there are attempts to use other kind of detectors such as e.g.  resistive plate chambers {\cite{6,7}}, straw tube drift chambers {\cite{8}} or plastic scintillators {\cite{9,10,11,12,13}}. The latter solution is developed in the framework of the J-PET collaboration aiming at the development of the low-price whole body scanner.  We are developing PET prototype with the axial alignment of plastic scintillator strips read out at two sides by photomultipliers {\cite{14,15,16}}. Relatively low price of plastic scintillators and the possibility of their easy preparation in different shapes and sizes allows, in contrast to the current crystal-based PET scanners, for the constructing of the large diagnostic chambers in a cost-effective way. Concurrently to the elaboration of the dedicated electronics {\cite{17}}, data acquisition \cite{18}, data analysis \cite{19,20}, and image reconstruction methods {\cite{21,22}} we carry out research and development {\cite{23}} aiming at the elaboration of the new plastic scintillators which would allow to improve time and energy resolution of the J-PET detector. Recently in the patent application a novel plastic scintillator is described {\cite{24}} novelty of the concept lies in the application of the 2-(4-styrylphenyl)benzoxazole as a wavelength shifter. In this article a light yield of such scintillator (referred to as a J-PET scintillator) is investigated.

Majority of plastic scintillators are ternary systems consisting of three components:  
a polymeric matrix and two different fluorescent additives referred to as a primary and 
a secondary additive – a wavelength shifter. Polystyrene or polyvinyltoluene constitutes polymer base for most of the commercially available plastic scintillators [23] and there is 
a large number of substances that can be used as fluorescent additives, e.g. PPO 
(2,5-diphenyloxazole) or PTP (p-terphenyl) as primary additives and POPOP (5-phenyl-2-[4-(5-phenyl-1,3-oxazol-2-yl)phenyl]-1,3-oxazole as a wavelength shifter. The secondary additive - wavelength shifter - shifts the wavelength of the emitted light to the visible range that matches the spectral sensitivity of photomultiplier \cite{25}. 

The molecule of 2-(4-styrylphenyl)benzoxazole (Figure 1) which is used for the first time as 
a scintillator dopant contains benzoxazole group - typical for scintillator additives - and 
a stilbene group \cite{24}. Stilbene is widely known organic crystalline scintillator and 
2-(4-styrylphenyl)benzoxazole is also well known substance which was used e.g. as an optical brightening agent and emissive material in electroluminescent OLED diodes \cite{26,27}.

\begin{figure}[h!]
\begin{center}
\includegraphics [scale=0.30] {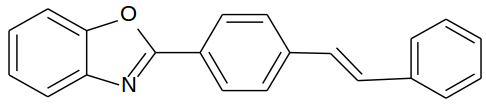}
\caption{Chemical structure of 2-(4-styrylphenyl)benzoxazole}
\end{center}
\end{figure}

In this article we will present that 2-(4-styrylphenyl)benzoxazole works also effectively as 
a wavelength shifter in plastic scintillators transferring light from the UV (330 - 380 nm) to the required visible range. These two ranges are adjusted respectively: the first one, to the emission wavelength of the primary fluor (PPO or PTP) and the second one to the photomultipliers spectral response (300-650 nm) \cite{28}. 

\section{Synthesis method}

J-PET plastic scintillator was produced by bulk polymerization in glass containers which were prepared before the polymerization by silanization treatment. The details of the method are described in Ref. [23]. Duration and temperature of the polymerization process were optimized and the solution of scintillator dopants in a liquid monomer was heated up to 140$^{\circ}C$. The polymerization process lasted about 100 hours in thermostatic furnace chamber and included: initial heating, heating, polymerization, annealing and final cooling to 30$^{\circ}C$. 
Because of the requirement of high optical properties, e.g. homogeneous dopant distribution, a~J-PET scintillator was prepared from a pure monomer.  The samples described in this article were prepared by dissolving in the styrene a 2 wt\% of 2,5-diphenylbenzoxazole as a first additive and 0.03~wt\% of 2-(4-styrylphenyl)benzoxazole as a wavelength shifter.
The synthesised J-PET scintillator was cut and polished to the cuboid shape with the dimensions of 14mm$\times$14mm$\times$20mm.

\section{Experimental setup}

The positron emission tomography is based on the registration of the annihilation gamma quanta with energy of 511 keV. Therefore, the light yield of new J-PET scintillator was tested by irradiation with the collimated beam of annihilation gamma quanta emitted by the 22Na isotope. The experiment was conducted by means of the setup shown in Figure 2. Two scintillators were placed in the setup: tested J-PET and reference BC-420 scintillator \cite{30} both cut to the cuboid shape with the same dimensions and wrapped with the 3M Vikuiti specular reflector foil \cite{29}. The light yield of the BC-420 plastic scintillator produced by Saint-Gobain amounts to 64\% of anthracene light output \cite{30}. This value is relatively high concerning the light yield of typical plastic scintillators which is within the range of 36\% - 68\% in comparison to anthracene light output \cite{30}.

\begin{figure}[h!]
\begin{center}
\includegraphics [scale=0.35] {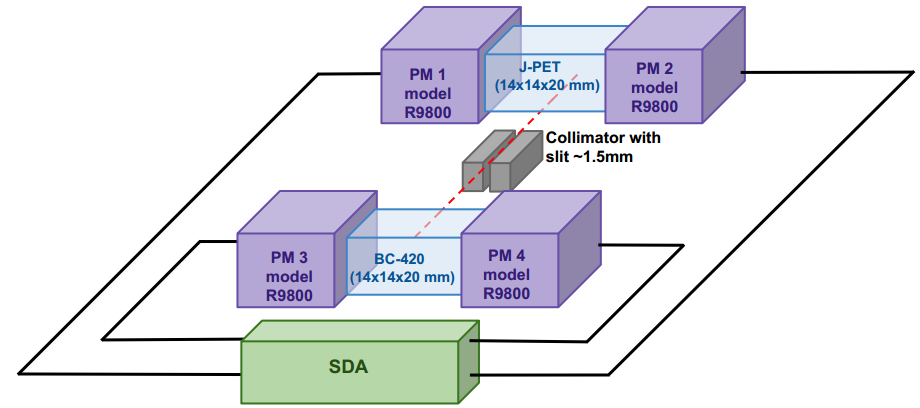}
\caption{A schematic view of the experimental setup used to compare the performance of the J-PET scintillator with respect to the scintillator BC-420. Detailed description is given in the text.}
\end{center}
\end{figure}

Each scintillator was connected at two sides by optical gel EJ-550 to Hamamatsu R9800 photomultipliers \cite{28}. The electric signals from photomultipliers were sampled by means of the Serial Data Analyzer (SDA). About 15000 events with coincident registration of signals from all four photomultipliers were collected. Subsequently the scintillators were dismounted and swapped with each other. Then the measurement was repeated. 

\section{Results}

In Figure 3 charge distributions of the registered signals for both measurements are shown. The charge of the signal is expressed in the number of photoelectrons estimated using 
a method described in reference \cite{31}. Plastic scintillators are composed of elements with low atomic number (carbon and hydrogen). Therefore, gamma quanta with the energy of 511 keV interacts in the plastic scintillator predominantly via Compton scattering leading to the continuous charge spectra as seen in Figure 3. 

The light output of the J-PET scintillator was calculated as a ratio between number of photoelectrons measured for the maximum energy deposition with J-PET and BC-420 scintillators multiplied by the known light yield of the BC-420 scintillator. In order to estimate a systematic uncertainty the measurement was performed two times replacing scintillators in the setup.  The estimated values of the light yield with respect to anthracene  amount to 58\% (left panel of Figure~3) and 62\% (right panel of  Figure~3). As a result we took an average value of 60\% $\pm$ 2\%, where the systematic error was estimated as a half of the difference between the obtained results.

\begin{figure}[h!]
\begin{center}
\includegraphics [scale=0.3] {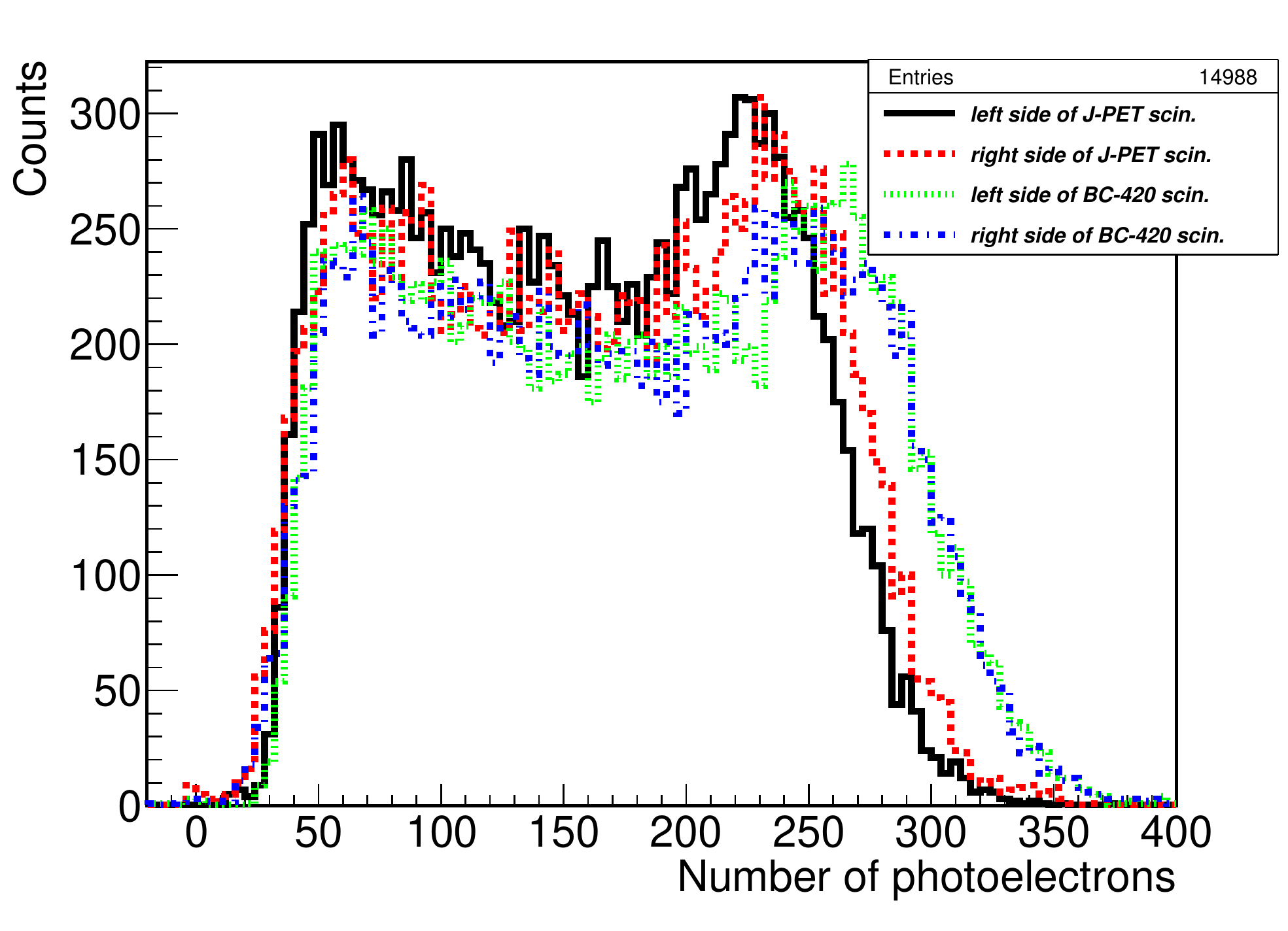} \includegraphics [scale=0.3] {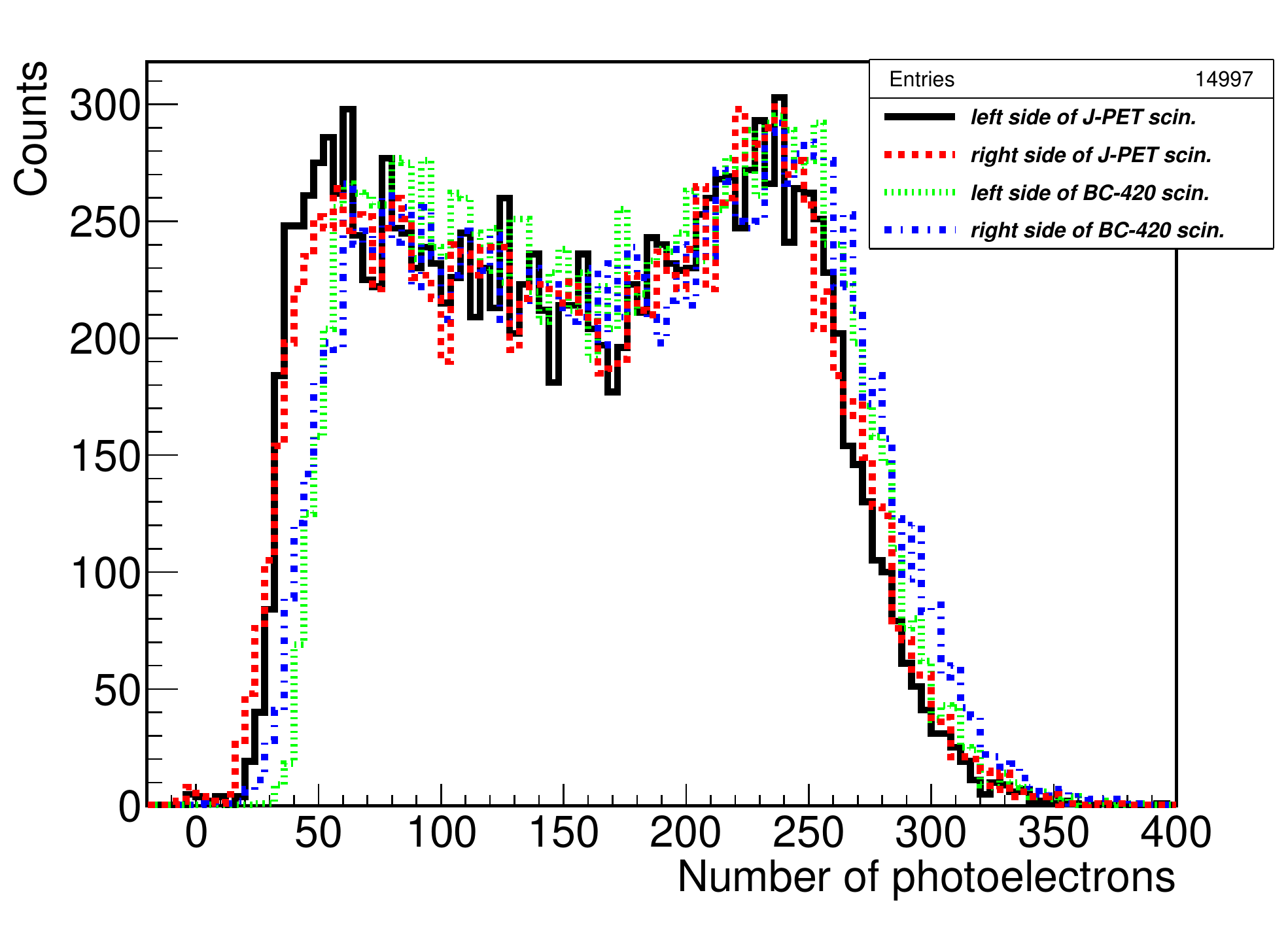}
\caption{Spectra of number of photoelectrons in signals from photomultipliers attached to the J-PET and 
BC-420 scintillator. Scintillators were irradiated at the centre by the collimated beam of 511 keV gamma quanta. The meaning of histograms is described in the legend. The left and right figures were obtained with the same experimental conditions except that the position of scintillators were replaced with each other. Left spectra were obtained with  J-PET scintillator connected to PM 1 and PM 2  and  BC-420 connected to PM 3 and PM 4, whereas in the right spectra results of the measurement are shown where J-PET scintillator is connected to PM 3 and PM 4 while BC-420 to PM 1 and PM 2.}
\end{center}
\end{figure} 

\section{Conclusions}

Light yield of the novel J-PET plastic scintillator synthesised from styrene doped with 2 wt\% of 2,5-diphenylbenzoxazole as a first additive and 0.03 wt\% of 2-(4-styrylphenyl)benzoxazole as a wavelength shifter was determined to be 60\% $\pm$ 2\%  of the anthracene light output. The synthesis of the J-PET scintillator with 2-(4-styrylphenyl)benzoxazole as a wavelength shifter is relatively simple and cost-effective. The obtained result for the light yield is comparable to the light outputs of best commercially available plastic scintillators \cite{30}. 
In this article we presented results of the pilot studies for which the concentration of the wavelength shifter has not been optimized. Therefore, there is still a potential to obtain higher light yields of the J-PET scintillator by optimization of the concentration of first and second additives.

\section{Acknowledgements}

We acknowledge technical and administrative support by M. Adamczyk, T. Gucwa-Ryś, 
A. Heczko, M. Kajetanowicz, G. Konopka-Cupiał, J. Majewski, W. Migdał, A. Misiak and the financial support by the Polish National Center for Development and Research through grant INNOTECH - K1/1N1/64/ 159174/NCBR/12, the Foundation for Polish Science through MPD programme and the EU and MSHE Grant No. POIG.02.03.00 - 161 00-013/09. The work on part of Institute of Chemistry and Physics was supported by DS 3133.

%

\end{document}